\documentclass[preprint,12pt]{elsarticle}

\usepackage{graphicx}

\usepackage{amssymb}
\usepackage{amsthm}
\usepackage{bm}
\def\Vec#1{\mbox{\boldmath $#1$}}
\def\eqne{\end{equation}}
 
\def\eqnb{\begin{equation}}
\def\PTP{Prog. Theor. Phys.(Kyoto)}
\def\NPA{{Nucl. Phys.} {\bf A}}
\def\NPB{{Nucl. Phys.} {\bf B}}
\def\PLB{{Phys. Lett.} B}

\def\PRL{Phys. Rev. Lett.}
\def\PRD{{Phys. Rev.} D}

\journal{Annals of Physics}

\begin{document}

\begin{frontmatter}



\title{Infrared fixed point of QCD, critical flavor number\\and triality automorphism of octonions}


\author{Sadataka Furui}

\address{School of Science and Engineering, Teikyo University.\\
1-1 Toyosatodai, Utsunomiya, 320-8551 Japan\thanks{\textit{E-mail address:} furui@umb.teikyo-u.ac.jp }}

\begin{abstract}
I show that the discrepancy on the critical flavor number of fermions $N_f^c$ for the appearance of the infrared fixed point based on the t'Hooft anomaly matching condition and derived from the Schr\"odinger functional method ($N_f^c\sim 9)$ and the experimental analysis of the JLab group using Bjorken sum rule and GDH sum rule, and our lattice simulation($N_f^c\sim 3$) could be resolved by assuming the topological structure of the infrared fixed point is not that of $U(1)^3$ but that of $G_2$ with triality automorphism of octonions which appear in the product of quaternions.

The agreement of the infrared fixed point of the running coupling measured in lattice simulations with use of the quaternion real condition and the prediction of the BLM renormalization theory might be due to the boundary condition on $S^3\times R$ manifold of quaternion.  The form factor of a proton measured via Ward identity through the difference of the inverse propagator at momentum p+q/2 and at p-q/2 agrees with the phenomenological dipole fit.

\end{abstract}


\end{frontmatter}


\section{Introduction}
\label{sect1}

The effective number of flavors of fermions in QCD is important in two aspects.
The first is the anomaly cancellation, i.e. since axial vector current Ward identity is not automatic, it is necessary to seek an anomaly cancellation mechanism to assure  the renormalizability of the field theory. The second is from the hadron phenomenology based on the perturbative QCD(pQCD). Based on the quenched lattice simulation data, Banks and Zaks\citep{BZ82} presented a conjecture, that when the energy becomes low and the QCD effective coupling becomes large, the beta function crosses a zero and an infrared fixed point appears when infrared singularity is suppressed by large number of quark pair creations.  The flavour number above which a fixed point appears is called critical flavor number $N_c$. The search of $N_c$ in lattice simulation was performed in Schr\"odinger functional method\citep{LNWW92} and Appelquist et al\citep{AFN08} claimed that it is around 10.

Although running coupling below QCD scale $\Lambda_{QCD}$ is difficult to interpret in pQCD, the running coupling at high energy can in principle extended to low energy via effective charge method\citep{Gr98, BP09}, and how to interplet the critical flavor number obtained in Schr\"odinger functional method is important. 

One can calculate effective coupling in lattice and in Dyson-Schwinger equation(DSE), via ghost-gluon coupling and via quark-gluon coupling. In the lattice simulation of the ghost-gluon coupling, we observed strong fluctuation of the ghost propagator and slight suppression of the infrared effective coupling.  
 In the DSE approach of quenched QCD, the effective ghost-gluon vertex in Landau gauge is calculated as, 
\[
\alpha(p^2)=\alpha(\mu^2)\frac{G(p^2,\mu^2)^2\tilde Z(p^2,\mu^2)}{G(\mu^2,\mu^2)^2\tilde Z(\mu^2,\mu^2)}
\]
where $D^{-1}(p^2)=\frac{p^2}{Z(p^2)}=\frac{1}{\tilde Z(p^2)}(p^2+m^2)$.

The term $m^2$ suppresses the fluctuation and the 'scaling solution', i.e. assume that the gluon propagator $z(p^2)\sim (p^2)^{-\kappa_A}$ and the ghost propagator $G(p^2)\sim (p^2)^{-\kappa_C}$ with $\kappa=\kappa_C=-\kappa_A/2$ yields the running coupling which is not infrared suppressed\citep{FMP09}. Although the qualitative behavior of the momentum dependence of the effective coupling of DSE and lattice simulation\citep{FN04} agree, I think the suppression of fluctuation via mass the term is artificial, and that the fluctuation of the ghost propagator is one of causes of the suppression of the infrared divergence of the ghost-gluon coupling. The DSE approach suffers from momentum cut-off, which could circumvented through the Pinch Technique\citep{BP09}.

The effective quark-gluon coupling in Coulomb gauge showed clear signal of infrared fixed point, and the result is compatible with the experimental values extracted from the polarized electron-proton scattering data \citep{DBCK06,DBCK08} via Bjorken sum rule\citep{Bj66} and Drell-Hearn-Gerasimov sum rule \citep{DH66, Ge65}.  The data of JLab group suggest infrared fixed point of $\alpha_s(0)\sim 3$.  From their fitted function of the effective coupling, the beta function obtained from the fitted line does not cross zero.  The similar beta function was obtained also in the ADS/QCD approach\citep{BdT10}.

In the infrared region, color confinement is a characteristic feature and we measured Kugo-Ojima parameter \citep{KO79} in quenched \citep{NF00} and unquenched gauge configurations\citep{FN06,sf08b}.  We observed that the Kugo-Ojima criterion is consistent with unquenched lattice simulation but about 20\% deviation occurs in quenched simulation. This qualitative difference suggests strong correlation between gauge field and the quark field.  Violation of the BRST symmetry in quenched calculation is observed also in DSE approach\citep{Dudal09,Kondo09,Kondo09b,FMP09}.
In \citep{sf08b}, I showed reduction of fluctuation in effective coupling and effective mass of quarks by choosing a gauge that the quarks on domain walls are correlated by self-dual gauge field (instanton).

 In the book of \'E. Cartan \citep{Cartan66}, a vector field like gauge field is produced from spinor fields as Pl\"ucker coodinates and the spinor is expressed in quaternion basis. I analyze in this paper, the lattice results of the Coulomb gauge quark propagator using quaternion basis and study consequences of the topological structure of the manifold. A product of quaternion makes an octonion and the octonion has the symmetry of exceptional Lie group $G_2$. It has the symmetry, which is called triality. It has the effect of triplicate flavor numbers.
Although in real world, quarks are massive and the system is not conformal, I expect proximity of our system to the conformal window.

In the infrared, when the Compton wave length of a quark $\displaystyle \frac{hc}{mc^2}$ where $m$ is the effective mass of a quark is large ($\sim 0.3GeV$), the proximity of other quarks in a baryon will change the boundary condition of quarks. I study the charge form factor of a proton using quarks in quaternion basis and modifying boundary condition of quarks. It makes a correction like final stated interaction and make the result gauge dependent.

This work is organized as follows. In sect.2, I summarize lattice data related to the IR fluctuation, and in sect.3, I summarize the Schr\"odinger functional method. In sect.4 anomaly matching condition is reviewed and in sect.5 the octonion and anomaly matching condition are explained.  I explain our method of gauge fixing of the gluon and domain wall fermion system in sect.6. The form factor of the proton calculated by this gauge fixing is presented in sect.7.

I discuss whether our finding of infrared fixed point in $N_f=2+1$ lattice simulation is incompatible with the estimates of Schr\"odinger functional method and give conclusions in sect.8.

\section{Unquenched lattice simulations and QCD effective coupling}

We performed lattice simulation of the effective coupling in Landau gauge from ghost-gluon coupling and found that it is close to the decoupling solution of the DSE. We found that the fluctuation of the ghost propagator in the infrared is large and the deviation of the Kugo-Ojima parameter in quenched simulation from the value expected from the BRST symmetry is due probably to this fluctuation. While in unquenched simulation, the fluctuation is suppressed and Kugo-Ojima parameter turned out to be consistent with the theory, and I guess the fluctuation is an essential property of the IR QCD.

In the case of effective coupling, we calculated the product of the ghost propagator and the gluon propagator and observed infrared suppression due to suppression of the ghost propagator. The ghost propagator in the Landau gauge has color antisymmetric part whose expectation value is 0 but its norm becomes comparable to that of color diagonal part in the infrared \citep{sf08a}. This kind of problem is absent in the Coulomb gauge. The lattice data calculated by using the gauge configuration of MILC \citep{MILC01} and  $16^3\times 32\times 16$ domain wall fermion of RBC/UKQCD \citep{AABB07} suggest an infrared fixed point of $\alpha_s(0)\simeq 3$ in the case of Coulomb gauge\citep{sf08b}.

The formula of Brodsky-Lepage-Mackenzie(BLM) scheme\citep{BLM83, BL95} strongly suggests a presence of infrared fixed point of $\alpha_s(0)=\pi$ and the effective coupling has walking like behavior near infrared.  
In the BLM scheme, observables A and B are related in terms of the effective charge as
\begin{equation}
\alpha_A(Q_A)=\alpha_B(Q_B)(1+r_{A/B}\frac{\alpha_B}{\pi}+\cdots)
\end{equation}
and the coefficient $r_{A/B}$ is so chosen that the result does not depend on the number of flavors $N_f$.

\begin{figure}\begin{minipage}[b]{0.47\linewidth}
\begin{center}
\includegraphics[width=6cm,angle=0,clip]{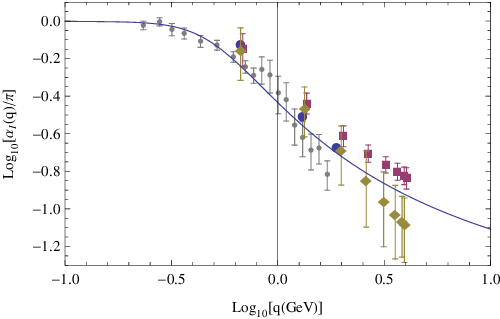}
\caption{The running coupling of the domain wall fermion. Coulomb gauge gluon-ghost coupling of $m_u=0.01/a{\rm (square)}$, $0.02/a{\rm (diamond)}$, and quark-gluon coupling of $m_u=0.01/a{\rm (large\, disks)}$. Small disks are the $\alpha_{s,g_1}$ derived from the spin structure function of the JLab group\citep{DBCK08}and the solid curve is their fit. (color online)}
\label{alp_plt}
\end{center}
\end{minipage}
\hfill
\begin{minipage}[b]{0.47\linewidth}
\begin{center}
\includegraphics[width=6cm,angle=0,clip]{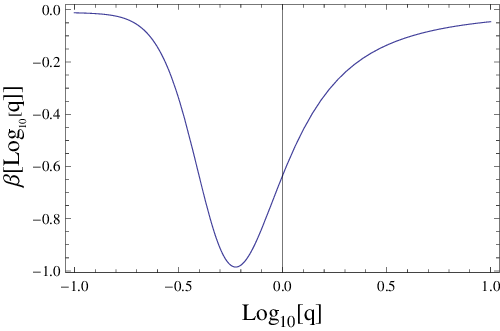}
\caption{The beta function calculated from the logarithmic derivative of the running coupling parametrized by the JLab group
.}
\end{center}
\end{minipage}
\end{figure}

I show in Fig. \ref{alp_plt}, the effective coupling of JLab experiment, and that of the lattice simulation of the quark gluon coupling.

In our lattice simulation of $N_f=2+1$, $\alpha_s(q)\sim 1.7$ at $q\sim 0.6$GeV and the deviation of the running coupling from 2-loop perturbative calculation is significant below $q\sim 3$GeV due to $A^2$ condensates \citep{Orsay02,KMSI02}. I think the non-perturbative effect is significant in the infrared and the Schr\"odinger functional method do not disprove the infrared fixed point at $N_f=2+1$.  In \citep{BS08,BT08}, the origin of appearance of the infrared fixed point is attributed to the acquiring of mass of the gluon and decoupling of the gluon polarization effects.

\section{Schr\"odinger functional and the critical flavor number}
Recently Appelquist\citep{App08,AFN08} claimed by performing lattice simulation of running coupling in the Schr\"odinger functional method \citep{LNWW92,LSWW94} that there is a critical number of flavors ${N_f}^c$ below which both chiral symmetry breaking and confinement set in. He assigned $8\leq N_f^c\leq 12$ and in the case of $N_f=8$ the running coupling monotonically increase as $\beta$ decreases and in the case of $16^4$ lattice and $\beta\sim 4.65$, $g^2(L)/4\pi\sim 20/4\pi\sim 1.7$, and that there is no sign of infrared fixed point. However, the Schr\"odinger functional is derived through the field that satisfies a boundary condition at $x_0=0$ and $x_0=T$
\begin{equation}
B_0=0, \quad B_k=[x^0 C_k'+(T-x^0)C_k]/T
\end{equation}
where in the case of SU(3)
where 
\begin{equation}
C_k=\frac{i}{L}\left(\begin{array}{ccc}
                           \phi_{k1}&0 &0\\
                           0&\phi_{k2} & 0\\
                           0 & 0&\phi_{k3}\end{array}\right)\quad{\rm and}
C_k'=\frac{i}{L}\left(\begin{array}{ccc}
                           \phi_{k1'}&0 &0\\
                           0&\phi_{k2'} & 0\\
                           0 & 0&\phi_{k3'}\end{array}\right)
\end{equation}
in which conditions ${\sum_i}^N_c\phi_{ki}=0$ and ${\sum_i}^N_c\phi_{ki'}=0$ for all $k$ is adopted.  

The functions $\phi$ and $\phi'$ are parametrized by $\eta$ and $\nu$ and the running coupling is derived from
\begin{equation}
\frac{\partial\Gamma}{\partial\eta}|_{\eta=\nu=0}=\frac{k}{\bar g^2}.
\end{equation}

The Schr\"odinger functional method $\alpha_s(q)$ is parametrized as $\displaystyle \frac{c}{\log q/\Lambda}+\cdots$ and its applicability in the region of $q$ below and around $\Lambda\sim 213\pm 40$MeV is not clear, since the vector field could be defined from fermionic field and the structure of the manifold above and below $\Lambda$ could be different. 

Brodsky et al, \citep{BS08,BT08} argue that in the DSE  $q<\Lambda$ or $r>1$fm, and in ADS/QCD $\zeta=\sqrt{{\Vec b}_\perp x(1-x)}>\Lambda_{max}\sim 1$fm, confinement is essential and the quark-gluon should not be treated as free. 

The QCD coupling constant is related to the bare coupling as
\begin{equation}
g={Z_A}^{3/2}Z_g^{-1}g_0
\end{equation}
and in one loop perturbation theory, the gluon self energy yields
\begin{equation}
Z_A=1+\frac{g_0^2}{16\pi^2}(\frac{13}{3}-\xi)t_2(V)\log\Lambda/\mu
\end{equation}
and the vertex renormalization factor is
\begin{equation}
Z_g=1+\frac{g_0^2}{16\pi^2}\left(\frac{17}{6}-\frac{3\xi}{2}\right)t_2(V)\log\Lambda/\mu
\end{equation}
where $\xi=1$ in Coulomb gauge and 
\begin{equation}
t_2(V)\delta^{ab}=C^{acd}C^{bcd}
\end{equation}
which is $n\delta^{ab}$ for $SU(n)$. Our conjecture is that, in the infrared region, the vector potential is produced as Pl\"ucker coordinate of fermions and that triality symmetry of the fermion effectively triplicate $t_2(V)$. 

\section{Anomaly matching condition and number of flavors}

Infrared QCD is characterized by spontaneous chiral symmetry breaking. The quark field of QCD at low energy is expressed as\citep{tH80}
\begin{equation}
\psi=\frac{1}{2}(1+\gamma_5)\psi_L+\frac{1}{2}(1-\gamma_5)\psi_R
\end{equation}
where $\psi_L$ transforms as $SU(3)$ color under $G_c$, $SU(N)_L$ flavor under $G_F$ and $SL(2,c)$ Lorentz spinor and $\psi_R$ transforms as $SU(3)$ color, $SU(2)_R$ flavor and $SL(2,c)$ Lorentz spinor.

Color gauge fields bind these quarks into baryons which must be color singlets. In the model that gauge group $G_c$ coupled to chiral fermion, the representation of the group must be such that the anomaly is canceled.  In the space of symmetry $G_c\times G_F$, t'Hooft considered
\begin{equation}
G_F=SU(n_1)_L\otimes SU(n_2)_R\otimes SU(n_3)_L\otimes SU(n_4)_R\otimes U(1)^3
\end{equation}
where $n_{1,2}$ refer to the triplets and $n_{3,4}$ to sextets.
He proposed anomaly matching relation in which the symmetric anomaly coefficient $d^{abc}$ contributed from the massless quarks should be equal to the coefficient $D^{abc}$ contributed from the massless color singlet composite fermions \citep{tH80}.

The approach up to fermionic degrees of freedom $N_f=5$ was not successful, but
an algebraic research was extensively done in \citep{Alb81}, and it is shown that the matching of $SU(N_f)_L^2 \times U(1)_B$ and $SU(N_f)_R^2 \times U(1)_B$ anomalies, where $U_B(1)$ is the vector-like Baryon number, and matching of $SU(N_f)_L^3$ and $SU(N_f)_R^3$ anomalies can be done in $N_f\pm 6$ dimensional representation \citep{Ter00}. 
The condition on $SU(N_f)^3_L$ and $SU(N_f)^3_R$ anomalies is
\begin{equation}
3=-(N_f\pm 6)+\frac{1}{2}(N_f\pm 3)(N_f\pm 6)+\frac{1}{2}N_f(N_f\pm 1)-N_f(N_f\pm 4)
\end{equation} 
and that for $SU(N_f)^2_L U_B(1)$ and $SU(N_f)^2_R U_B(1)$ anomalies is
\begin{equation}
1=-(N_f\pm 6)\frac{N_f\pm 2}{N_f\pm 6}+\frac{1}{2}(N_f\pm 2)(N_f\pm 3)+\frac{1}{2}N_f(N_f\pm 1)-N_f(N_f\pm 2).
\end{equation}
where $U_B(1)$ is the baryon number charge.
It means that QCD with 3 flavor and 3 color system cannot satisfy the matching condition. 

Sannino \citep{San09} found another solution that satisfies the anomaly matching condition in $2N_f\pm 15$ dimensional system, which requires $N_f>8$. 
He found the condition for a $SU(N_f)_L^3$ anomalies as
\begin{equation}
3=-(2N_f\pm 15)+\frac{1}{2}(N_f\pm 3)(N_f\pm 6)+\frac{1}{2}N_f(N_f\pm 1)-N_f(N_f\pm 4)
\end{equation} 
and that for  $SU(N_f)_{L}^2 U_V(1)$ anomalies as
\begin{equation}
1=-(2N_f\pm 15)\frac{2N_f\pm 5}{2N_f\pm 15}+\frac{1}{2}(N_f\pm 2)(N_f\pm 3)+\frac{1}{2}N_f(N_f\pm 1)-N_f(N_f\pm 2)
\end{equation}
where $U_V(1)$ is the vector charge.

The critical number of flavors for opning the conformal window\citep{Gr98} and the critical number of flavors for presence of infrared fixed point is expected to be close to each other. A study of infrared fixed point was done on the lattice by measuring the momentum dependence of the QCD running coupling.

In the perturbative $\beta$ function and DSE approach, infrared fixed point $\alpha_{IR}$ is a decreasing function of number of flavors $N_f$. When $\alpha_{IR}$ decreases below the minimal value of $\alpha_{cr}$ for which the fermion acquires a mass, the $N_f$ is called ${N_f}^c$\citep{ATW96,BS08}.  Near this $N_f$, the effective coupling varies slowly as the momentum changes. Lattice simulation of $N_c=3$\citep{JLQCD92} suggests ${N_f}^c\sim 7$, while more recent lattice simulation \citep{App08,AFN08,AFN08e} and DSE\citep{ATW96} suggest that $8<{N_f}^c\leq 12$. 

\section{Octonion and Anomaly matching condition}

In the vector like theory that $SU(N_f)_L^2 \times U(1)_V$ and $SU(N_f)_R^2\times U(1)_V$ anomalies and $SU(N_f)_L^3$ and $SU(N_f)_R^3$ anomalies can be matched in $2N_f\pm 5N_c$ dimensional representation\citep{San09}.
The model of \citep{San09} may be regarded as magnetic dual of the gauge theory.
The factor 5 in this QCD dual may be related to the $1/N_c$ corrections to the one body matrix elements in the Skyrme model\citep{ABO86} which was necessary to reproduce the magnetic transition of a nucleon to $\Delta$.
These condition $N_f-6=3$ and $2N_f-15=3$ are both satisfied by $N_f=9$, but the standard model of QCD predicts $N_f=3$.  
The quadratic form of quaternion is expressed by octonion, and the automorphism in the space of octonion is $G_2$ group which is exceptional Lie group with 14 dimensional representation. In $G_2$, there is a specific automorphism, which is called triality. In the following subsection, I study the structure of the $G_2$ group.

\subsection{Representation theory of $G_2$ group}
In this subsection, I study the quaternion and the 
representation theory of $G_2$\citep{FH04}.

A quaternion $q\epsilon \bf H$ can be defined by using the basis $\{1,{\bf i},{\bf j},{\bf k}\}=\{I,i\sigma_1,i\sigma_2,i\sigma_3\}$ as $q=wI+ix\sigma_1+iy\sigma_2+iz\sigma_3$, where $\sigma_i$ are Pauli matrices and $I =1^2$ is the unit matrix. 
As elements of a complex number is expressed by a product of real numbers ${\mathcal C}={\mathcal R}+{\bf i}{\mathcal R}$:
\begin{equation}
(x_1,y_1)(x_2,y_2)=(x_1 x_2-y_1 y_2, x_1y_2+y_1 x_2)\nonumber
\end{equation}
a quaternion is given from bijection of two complex numbers
\[
\zeta_1=\left(\begin{array}{cc} z_1& w_1\\
                                 -\bar{w_1}&\bar{z_1}\end{array}\right),
\qquad \zeta_2=\left(\begin{array}{cc} z_2& w_2\\
                                  -\bar{w_2}&\bar{z_2}\end{array}\right)
\]
and the elements of quaternions are expressed by a product of complex numbers 
as the first low of the product of $\zeta_1$ and $\zeta_2$ shows:
\begin{equation}
(z_1, w_1)(z_2,w_2)=(z_1 z_2-w_1\bar w_2, z_1 w_2+w_1 \bar z_2)\nonumber
\end{equation}

Thus ${\mathcal H}={\mathcal C}+{\bf j}{\mathcal C}$ \citep{KR91}.
The quaternion $\bf H$ and a new imaginary unit ${\bf l}$ that anti-commute with ${\bf i}, {\bf j},{\bf k}$ compose an octonion ${\mathcal O}={\mathcal H}+{\bf l} {\mathcal H}$. It is spanned by
\begin{equation}
\{ 1,{\bf i},{\bf j},{\bf k},{\bf l},{\bf i}{\bf l},{\bf j}{\bf l},{\bf k}{\bf l} \}=\{1,{\bf e_1, e_2, e_3, e_4, e_5, e_6, e_7}\}\nonumber
\end{equation}
with bases of Clifford algebra ${\bf O}={\bf R}+{\bf R}^7$.
 
Elements of an octonion is expressed as a product of quaternions
\begin{equation}
(p_1,q_1)\circ(p_2,q_2)=(p_1p_2-\bar q_2 q_1, q_2 p_1+q_1\bar p_2)
\end{equation} 

The algebra ${\bf O}$ is associated with an involution\citep{Port95}
\begin{equation}
g\to \check g=\left(\begin{array}{cc} 1&0\\
                                      0&-^7 1\end{array}\right)g
              \left(\begin{array}{cc} 1&0\\
                                      0&-^7 1\end{array}\right) 
\end{equation}
where $^7 1$ is the 7 dimensional diagonal matrix.
 In the subspace ${\bf R}^7$, automorphisms of $SO(7)$ are of the form $U\to SUS^{-1}$ where $S\epsilon SO(7)$. Automorphism of ${\bf Spin}(7)$ are of the form $u\to sus^{-1}$. 

In the space ${\bf R}^8$, automorphism of ${Spin}(8)$ is expressed as injection 
\begin{equation}
{\bf R}^8\to {^2{\bf R}}(8): a\left(\begin{array}{cc}\upsilon(a)& 0\\
                                                  0 & ^t\upsilon(a)\end{array}\right)
\end{equation}
whose image is defined as $\bf Y$. For an element $y\in \bf Y$ and the unit element $g\in{\bf O}$, involution $\check g$ of $g$ is defined as
\begin{equation}
\check g y {\bf e}=\overline{ g {\overline {y {\bf e}}} }.
\end{equation}
where $\overline {y{\bf e}}$ is the complex conjugation of $y{\bf e}$.
When the element $g$ satisfies $g {\bf e}={\bf e}$ and $\check g$ also satisfies $\check g {\bf e}={\bf e}$, $\check g$ is called the companion element of $g$.

In the case of SO(8), with $g_0$ and the companion of $g_1$, an element of $Spin(8)$ is
 $\left( \begin{array}{cc} g_0& 0\\
                           0&\check g_1\end{array}\right)$
and by the definition of $g_2$ of
\begin{equation}
g_0 y \check g_1^{-1}{\bf e}=\check g_2 y {\bf e}
\end{equation}
for all $y\in {\bf Y}$, one can construct a triple ($g_0,g_1,g_2$).

The triality automorphism is
\begin{equation}
\theta: Spin(8)\to Spin(8); \left(\begin{array}{cc} g_0& 0\\
                                                  0&\check g_1\end{array}\right)\to\left(\begin{array}{cc} g_1&0\\
                           0&\check g_2\end{array}\right)
\end{equation}

Triality transformation rotates 24 dimensional bases defined by Cartan\citep{Cartan66}.

\[
\{\xi_0, \xi_1, \xi_2, \xi_3, \xi_4\},\quad
\{\xi_{12}, \xi_{31}, \xi_{23}, \xi_{14}, \xi_{24}, \xi_{34}\},\quad
\{\xi_{123}, \xi_{124}, \xi_{314}, \xi_{234}, \xi_{1234}\}
\]
\[
\{x^1, x^2, x^3, x^4\}, \quad \{x^{1'}, x^{2'}, x^{3'}, x^{4'}\}
\]

The trilinear form in these bases is
\begin{eqnarray}
{\mathcal F}&=&\phi^TCX\psi=x^1(\xi_{12}\xi_{314}-\xi_{31}\xi_{124}-\xi_{14}\xi_{123}+\xi_{1234}\xi_1)\nonumber\\
&+&x^2(\xi_{23}\xi_{124}-\xi_{12}\xi_{234}-\xi_{24}\xi_{123}+\xi_{1234}\xi_2)\nonumber\\
&+&x^3(\xi_{31}\xi_{234}-\xi_{23}\xi_{314}-\xi_{34}\xi_{123}+\xi_{1234}\xi_3)\nonumber\\
&+&x^4(-\xi_{14}\xi_{234}-\xi_{24}\xi_{314}-\xi_{34}\xi_{124}+\xi_{1234}\xi_4)\nonumber\\
&+&x^{1'}(-\xi_{0}\xi_{234}+\xi_{23}\xi_{4}-\xi_{24}\xi_{3}+\xi_{34}\xi_2)\nonumber\\
&+&x^{2'}(-\xi_{0}\xi_{314}+\xi_{31}\xi_{4}-\xi_{34}\xi_{1}+\xi_{14}\xi_3)\nonumber\\
&+&x^{3'}(-\xi_{0}\xi_{124}+\xi_{12}\xi_{4}-\xi_{14}\xi_{2}+\xi_{24}\xi_1)\nonumber\\
&+&x^{4'}(\xi_{0}\xi_{123}-\xi_{23}\xi_{1}-\xi_{31}\xi_{2}-\xi_{12}\xi_3)
\end{eqnarray}
There are three semi-spinors which have a quadratic form which is invariant with respect to the group of rotation
\begin{eqnarray}
\Phi=^t\phi C\phi&=&\xi_0\xi_{1234}-\xi_{23}\xi_{14}-\xi_{31}\xi_{24}-\xi_{12}\xi_{34}\\
\Psi=^t\psi C\psi&=&\xi_1\xi_{234}-\xi_{2}\xi_{134}-\xi_{3}\xi_{124}-\xi_{4}\xi_{123}\end{eqnarray}
and the vector
\begin{equation}
F=x^1 x^{1'}+x^2 x^{2'}+x^3 x^{3'}+x^4 x^{4'}
\end{equation}

The triality transformation that makes $g{\bf e} ={\bf e}$ for ${\bf e}=x^1,x^2,x^3$ and $x^{1'},x^{2'},x^{3'}$ is defined as $G_{23}$, $g{\bf e} ={\bf e}$ for 
${\bf e}=\xi_1,\xi_2,\xi_3$ and $\xi_{124},\xi_{314},\xi_{234}$ is defined as $G_{12}$, $g{\bf e} ={\bf e}$ for ${\bf e}=\xi_{12},\xi_{31},\xi_{23}$ and $\xi_{14},\xi_{24},\xi_{34}$ is defined as $G_{13}$. In additioin to $G_{23}, G_{12}$ and $G_{13}$, there are automorphism ${^t(}G_{13}G_{12})=G_{132}$ and ${^t(}G_{12}G_{13})=G_{123}$. 

When one defines for each $i=0,1,2$
\[
H_i=\{(g_0,g_1,g_2)\in Spin(8): g_i {\bf e}={\bf e}\},
\]
the triality transformation is expressed as $\theta(H_0)=H_1, \theta(H_1)=H_2$ and $\theta(H_2)=H_0$,  and   $G_2=H_1\cap H_2=H_2\cap H_0=H_0\cap H_1$\citep{Port95}.

In the space of octonions, I choose $\xi_1$ as an element orthogonal to 1, and $\xi_2$ as an element orthogonal to 1 and $\xi_1$, and $\xi_4$ as an element orthogonal to 1, $\xi_1, \xi_2$ and $\xi_1\xi_2$. For orthogonal elements $\xi_1, \xi_2, \xi_4$ chosen as $\xi_k$, there is an automorphism $\xi_1\to \xi_{14}, \xi_2\to \xi_{24}, \xi_4\to \xi_0$.
The choice of $\xi_{14}$ is a choice of a point on a unit sphere in the 7 dimensional space, $\xi_{24}$ is a choice of a point on a unit sphere in the 6 dimensional space, and $\xi_0$ is a choice in the space orthogonal to $1, \xi_{14}, \xi_{24}, \xi_{14}\xi_{24}$ i.e. in the 4 dimensional space. These form 6, 5 and 3 dimensional manifolds. 
The exceptional Lie group representation $G_2$ is the automorphism in this 14 dimensional manifold.

I assign $SU(3)$ fundamental color representation of quarks on the 3 dimensional manifolds and that of the color representation of diquarks on the 6 dimensional manifolds and 4 dimensional spinors and  1 dimensional space between spins on the domain walls to the 5 dimensional manifold. The triality transformation mixes the 3 and 6 dimensional spaces.

  Irreducible representations of $g_2$ algebra with highest weight $a\omega_1+b\omega_2$ is written as $\Gamma_{a,b}$, and I study the standard representation $V=\Gamma_{1,0}$.  

\subsection{Anomaly matching condition}

A problem is whether the anomaly matching of the color triplet and color sextet sectors in a baryon can be explicitly realized. The fundamental representation of the $g_2$ algebra has 14 vectors
\begin{equation}
g_2=\{H_1,H_2,X_1,Y_1,X_2,Y_2,X_3,Y_3,X_4,Y_4,X_5,Y_5,X_6,Y_6\}\nonumber
\end{equation}
One can define a subgroup
\begin{equation}
g_0=\{H_5,H_2,X_5,Y_5,X_2,Y_2,X_6,Y_6\},\nonumber
\end{equation}
where $H_5=H_1+H_2$, which is isomorphic to $sl_3(C)$\citep{FH04}.

The rest of the Lie algebra consists of
\begin{equation}
W=\{X_4,Y_1,Y_3\}\quad {\, \rm and }\quad W^*=\{Y_4,X_1,X_3\}\nonumber
\end{equation}

When a representation has the highest weight $\omega_1=2\alpha_1+\alpha_2$ it is seven dimensional, we call it $V$ . The wedge product $\wedge^2 V$ is
\begin{equation}
\wedge^2 V\simeq \Gamma_{0,1}+V\nonumber
\end{equation}
and $\wedge^3 V$ is
\begin{equation}
\wedge^3 V\simeq \Gamma_{2,0}+V+{\bf C}\nonumber
\end{equation}
where {\bf C} is the trivial representation.

The action of $g_2$ on the standard representation preserves a skew-symmetric trilinear form $\omega$ on $V$, which consists of
\begin{equation}
\omega=w_3\wedge u\wedge v_3+v_4\wedge u\wedge w_4+w_1\wedge u\wedge v_1\\+2v_1\wedge v_3\wedge w_4+2w_1\wedge w_3\wedge v_4
\end{equation}
where with a use of root vectors $Y_1$ and $Y_2$, $v_3=Y_1(v_4), v_1=-Y_2(v_3), u=Y_1(v_1), w_1=Y_1(u)/2, w_3=Y_2(w_1)$ and $w 4=-Y_1(w 3)$.
It means that the five-dimensional space $\omega$ is stable in $\wedge^3 V$,
and that the anomaly matching relation of the massless fermion state and the bound state of three massless fermion which is given by $\wedge^3 V$ can be
considered in the same five dimensional space. 

$su(3)$ is a subalgebra of $sl_3(C)$, but it is a real representation. The quaternion real condition imposed on the quark propagator has the effect of constraining $sl_3(C)$ to its real representation $su(3)$. Anomaly matching relation on
the massless fermion and the bound state of three massless fermions with flavor $N_f=3$ and color $N_c=3$ could be interpreted as the $N_f=9$ without triality, since the same 5 dimensional space appears 3 times through triality transformation.  
The topology around the infrared fixed point would be more complicated than the $U(1)^3$.

\section{Gauge fixing of the domain wall fermion}

We first perform a gauge fixing in Landau gauge using the over relaxation method and then fix to the Coulomb gauge without touching the $A_4$ component\citep{NF05}. I then measure the quark propagator and fix the gauge in the space of the 5th dimension between the domain walls.

In the analysis of instantons in quaternion bases, Corrigan and Goddard \citep{CG81} defined the transition function $g(\omega,\pi)$ where $\omega$ and $\pi$ are complex two-spinor which satisfy
\[
g(\lambda\omega,\lambda\pi)=g(\omega,\pi), \quad det\, g=1.
\]
@When $\omega=x\pi$ where $x=x^0-i {\Vec x}\cdot{\Vec\sigma}$, $g(\omega,\pi)$ is a quaternion can be expressed as
\[
g(x\pi,\pi)=h(x,\zeta)k(x,\zeta)^{-1}
\]where $\zeta=\frac{\pi_1}{\pi_2}$, $h(x,\zeta)$ is regular in $|\zeta|>1-\epsilon$ and $k(x,\zeta)$ is regular in $|\zeta|<1+\epsilon$.

In \citep{CG81},  the Ansatz
\begin{eqnarray}
g_0&=&\left(\begin{array}{cc}e^{-\nu}&0\\
                           0&e^\nu\end{array}\right)
\left(\begin{array}{cc}\zeta^1&\rho\\
                       0&\zeta^{-1}\end{array}\right)
\left(\begin{array}{cc}e^{\mu}&0\\
                           0&e^{-\mu}\end{array}\right)\nonumber\\
&=&\left(\begin{array}{cc}e^\gamma\zeta^1&f(\gamma,\zeta)\\
                       0&e^{-\gamma}\zeta^{-1}\end{array}\right)
\end{eqnarray}
was proposed as the transformation matrix.  In our 5-dimesional domain wall fermion case, $\gamma=\mu-\nu$ and $\mu, \nu$ contain the phase in the 5th direction $i\eta$. 
\begin{equation}
2\mu=i\omega_2/\pi_2-i\eta=(x_1+ix_2)\zeta+ix_0-x_3-i\eta
\end{equation}
\begin{equation}
2\nu=i\omega_1/\pi_1+i\eta=(x_1-ix_2)\zeta+ix_0+x_3+i\eta
\end{equation}

 The quaternion reality condition of the transformation matrix $g(\gamma,\zeta)$ gives
\begin{equation}
\left(\begin{array}{cc}a_{L_s-1}& b_{L_s-1}\\
                       c_{L_s-1}& d_{L_s-1}\end{array}\right)
\left(\begin{array}{cc}\zeta^{1}e^\gamma& f\\
                       0&\zeta^{-1}e^{-\gamma}\end{array}\right)
=\left(\begin{array}{cc}\zeta^{1}e^{-\gamma}& \bar f\\
                       0&\zeta^{-1}e^{\gamma}\end{array}\right)
\left(\begin{array}{cc}a_{0}& b_{0}\\
                       c_{0}& d_{0}\end{array}\right)
\end{equation}
where $\displaystyle \bar f=\overline{f(\bar\gamma,-\frac{1}{\bar\zeta})}$.

The function $f$ and $\bar f$ taken in \citep{CG81} is
\[
f=\frac{d_0 e^\gamma-\frac{1}{a_{L_s-1}}e^{-\gamma}}{\psi},\quad 
\bar f=\frac{\frac{1}{d_{L_s-1}} e^\gamma-a_0e^{-\gamma}}{\psi}.
\]

In general $c_0$ and $c_{L_s-1}$ are polynomials of $\zeta$, and they satisfy $c_{L_s-1}\zeta^{1}=c_0 \zeta^{-1}$. I define
\[
\psi=\hat c_{-1}\zeta^{-1}+\hat c_{1}\zeta^{1}+\delta
\]
where $\hat c_{-1}=c_{L_s-1}$ and $\hat c_{1}=c_0$ and $\delta$ is a constant, which is defined later. 

With the Ansatz $\displaystyle \psi=c_0\zeta^{-1}+c_{L_s-1}\zeta^{1}+\delta$, I find $\displaystyle \zeta^{1}=\sqrt{\frac{c_0}{c_{L_s-1}}}$ and 
the $e^\gamma$ and the $\delta$ are calculated from the simultaneus equation (\ref{gamma}):
\begin{eqnarray}
\left\{\begin{array}{l}
a_{L_s-1}f+b_{L_s-1}\zeta^{-1}e^{-\gamma}=b_0\zeta^{-1}e^{-\gamma}+d_0 \bar f\nonumber\\
c_{L_s-1}f+d_{L_s-1}\zeta^{-1}e^{-\gamma}=d_0\zeta^{-1}e^{\gamma}.
\end{array}
\right.
\end{eqnarray}\label{gamma}
The equation has two sets of solutions, which can be obtained numerically. I assign  
$\displaystyle
\left(\begin{array}{cc}a_{0}& b_{0}\\
                       c_{0}& d_{0}\end{array}\right)_{L/R} {\rm and}
\left(\begin{array}{cc}a_{L_s-1}& b_{L_s-1}\\
                       c_{L_s-1}& d_{L_s-1}\end{array}\right)_{L/R}
$
from the color diagonal components of the lattice data.

From the numerical practice, I observe that the deviation $\Delta L/R$ becomes small when $|\delta|$ is large, i.e. when $\psi\sim\delta\sim \frac{1}{f}$ is large. It means that $\displaystyle e^{2\gamma}\sim\frac{d_{L_s-1}}{d_0}$ is a good approximation in the case of small deviation, and I select the gauge such that the corresponding $|\delta|$ is large.

\section{Form factor of the proton}
With the prescription on the gauge of fermion on the domain walls, I calculate the baryon charge form factor of a proton using the SU(6) spin-flavor wave function.
The nucleon three point function is
\begin{eqnarray}
&&\langle G^{NV_\mu N}(t_2,t_1;{\Vec p}+\frac{\Vec q}{2},{\Vec p}-\frac{\Vec q}{2};\Gamma)\rangle\nonumber\\
&&=\sum_{\Vec x}e^{-i({\Vec p}+\frac{\Vec q}{2})\cdot {\Vec x}_2} e^{i({\Vec p}-\frac{\Vec q}{2})\cdot{\Vec x}_1} 
\times\Gamma^{\beta\alpha} \langle \Omega |T[\chi^\alpha({\Vec x}_2,t_2)V_\mu({\Vec 0},0){\bar\chi}^\beta({\Vec x}_1,-t_1)]\Omega\rangle\nonumber
\end{eqnarray}
where
\[
\chi(x)=\epsilon^{abc}[u^{Ta}(x)C\gamma_5 d^b(x)]u^c(x)
\]
In these expressions $abc$ specify colors, $\alpha$ and $\beta$ specify a spin dependent $A$ term or a spin independent $B$ term of each three quark propagators.

In the limit of $t_1\to -\infty, t_2\to +\infty$, the fourier transform reduces
to the vertex sandwiched by the propagators.
\subsection{The case of $q=0$}
I define the normalization by the ratio of the three point function $G^{NV_\mu N}$ and the two-point function defined as
\begin{eqnarray}
&&\langle G^{NN}(t_2,t_1,{\Vec p};\Gamma)\rangle=\sum_{\Vec x}e^{-i{\Vec p}\cdot{\Vec x}_2} e^{i{\Vec p}\cdot {\Vec x}_1}
\Gamma^{\beta\alpha}\langle \Omega |T\chi^\alpha({\Vec x}_2,t_2){\bar\chi}^\beta({\Vec x}_1,t_1)|\Omega\rangle.\nonumber
\end{eqnarray}

I take the limit of  $t_1\to -\infty, t_2\to +\infty$, here also.
I define the propagator
\begin{equation}
S(a, h_a, a, h'_a|p)=S^A(a,h_a,a,h'_a|p)+S^B(a,h_a,a,h'_a|p),
\end{equation}
\noindent where
\[
S^A(a,h a,a,h'_a|p)=\frac{-i{\mathcal A}p}{{\mathcal A}p^2+{\mathcal M}{\mathcal B}}\left|_{h_a, h'_a}\right.
\]
\[
S^B(a,h_a,a,h'_a|p)=\frac{\mathcal B}{{\mathcal A}p^2+{\mathcal M}{\mathcal B}}\left|_{h_a, h'_a}\right.
\]

In the coupling of $\Lambda_\mu$ to $(u^a(x) C\gamma_5 d^b(x))u^c(x)$ and $(u^{a'}(x) C\gamma_5 d^{b'}(x))u^{c'}(x)$, I evaluate the real part of
\begin{enumerate}
\item  $S^\alpha(c,h_c, c,h'_c|p) {u^c}_{h'_c}(p)\Lambda_\mu {u^c}_{h"_c}(p)\times \gamma_5 S^\beta(c, h"_c,c, h_c|p)^\dagger\gamma_5$

\item  $S^\alpha(a,h_a, a,h'_a|p) {u^a}_{h'_a}(p)\Lambda_\mu {u^a}_{h"_a}(p)\times \gamma_5 S^\beta(a, h"_a,a, h_a|p)^\dagger\gamma_5$

\item  $S^\alpha(b,-{h_b}, b,-{h'_b}|-p)^\dagger {u^b}_{h'_b}(-p)\Lambda_\mu {u^b}_{h"_b}(-p) \times\gamma_5 S^\beta(b,-{h"_b},b,-{h_b}|-p)\gamma_5$
\end{enumerate}

I used the fact that the helicity of $C\gamma_5 d(x)$ becomes -1 times the original.

The expectation values of the charge form factor $G^{NVN}_{c+}$ is
\begin{eqnarray}
\Gamma(p,q=0)&=&\frac{1}{2}[\frac{{G^{NVN}}_{c+}(p,0)}{G^{NN}_{c+}(p,0)}+\frac{{G^{NVN}}_{c-}(p,0)}{G^{NN}_{c-}(p,0)}
+\frac{{G^{NVN}}_{a+}(p,0)}{G^{NN}_{a+}(p,0)}+\frac{{G^{NVN}}_{a-}(p,0)}{G^{NN}_{a-}(p,q)}\nonumber\\
&&+\frac{{G^{NVN}}_{b-}(p,0)}{{G^{NN}}_{b-}(p,0)}+\frac{{G^{NVN}}_{b+}(p,0)}{{G^{NN}}_{b+}(p,0)}]\nonumber
\end{eqnarray}
At zero momentum the left-handed quark contribution dominates when the gauge fixing parity on the
domain wall fermion is applied. When the gauge fixing parity is not applied, the left-handed quark
and the right handed quark give almost the same contribution.

\begin{figure}
\begin{center}
\includegraphics[width=6cm,angle=0,clip]{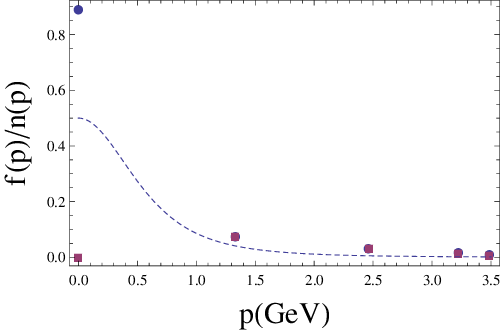}
\caption{The form factor of a proton using the DWF $m_u=0.01/a$ gauge fixed such that the fermion on the left wall and the right wall are correlated by self-dual gauge field, and the dipole fit with $M^2=(0.71$GeV$)^2$. At zero momentum left-handed(LH) fermion dominates the form factor, and at finite momentum, LH and RH contribute almost equally. (color online)}
\label{charge_plt01c}
\end{center}
\end{figure}

I show in Fig. \ref{charge_plt01c}, the form factor of the proton of the domain wall fermion of $16^3\times 32\times 16$ lattice of $ud-$ quark mass $0.01/a$ (148 samples). The momentum ${\bf p}$ is chosen to be $(p_1,p_2,p_3,p_4)=(0,0,0,0),(1,0,0,0),(1,1,1,0),(4,0,0,0),$
$(2,2,2,0),(3,3,3,0),(4,4,4,0)$. 

The dotted line is the dipole fit with $M^2=(0.71$GeV$)^2$. Since the contribution of left-handed and right-handed are almost the same for $p\ne 0$ the fitted
line is normalized to be 0.5 at $p=0$.
\subsection{The case of $q\ne 0$}
In \citep{MPSTV94}, the local vector current Ward identity on quark state is written as
\begin{equation}
q_\mu[Z_{V^L}\Lambda_{{V^L}_\mu}(p+\frac{q}{2},p-\frac{q}{2})]=-i(S^{-1}(p+\frac{q}{2})-S^{-1}(p-\frac{q}{2})\label{eqLamda}
\end{equation}
where $\displaystyle \Lambda_{{V^L}_\mu}(p+\frac{q}{2},p-\frac{q}{2})$ is the amputated vertex with momentum transfer $q$ and $Z_{V^L}$ is the renormalization factor. 
When one operates $\partial/\partial q_\rho$ on this expression, one obtains
\begin{equation}
Z_{V^L}(\Lambda_{V_\rho}^L(p)+q_\mu\frac{\partial}{\partial q_\rho}\Lambda_{V_\rho}^L(p+\frac{q}{2},p-\frac{q}{2})|_{q=0})=-i\frac{\partial}{\partial p_\rho}S^{-1}(p)
\end{equation}
In our lattice simulation, I take the difference of vertex of $\displaystyle p+\frac{q}{2}$  and $\displaystyle p-\frac{q}{2}$ in which $q$ is parallel to $p$, I consider the difference near the cylinder cut momenta $p=(\bar p,\bar p,\bar p,2\bar p), (\bar p=1,2,3,4)$, i.e. the difference of 
 $\displaystyle p-\frac{q}{2}=(1,2,2,4),(2,3,3,6)$ and $\displaystyle p+\frac{q}{2}=(3,2,2,4),(4,3,3,6)$, respectively in the case of $\rho=1$, and $\displaystyle p-\frac{q}{2}=(2,1,2,4),(3,2,3,6)$ and $\displaystyle p+\frac{q}{2}=(2,3,2,4),(3,4,3,6)$, respectively in the case of $\rho=2$ .

I calculate for each three body vertex $G^{NVN}_c(p,q)$ in $\displaystyle S^{-1}(p\pm \frac{q}{2})$ as
\begin{eqnarray}
&&\langle u^a(x) C\gamma_5 d^b(x))u^c(x)|S \gamma_4 \Lambda_\mu S^\dagger| (u^{a}(x) C\gamma_5 d^{b}(x))u^{c}(x)\rangle\nonumber\\
&&=S(c\sigma, c\sigma'|p+\frac{q}{2})[u^c\sigma' \Lambda_\mu u^c\sigma" ]\gamma_5 S(c\sigma", c\sigma|p+\frac{q}{2})^\dagger
\end{eqnarray}
Although gluon can change color of the quark, I select color diagonal component and spin off-diagonal components.

The nucleon helicity $\sigma$ is chosen to be $+$ and but at the vertex $\Lambda$, the helicity of the quark may change through the propagator$S$.

The two body part $\displaystyle G^{NN}(p\pm \frac{q}{2})$ is evaluated by
making a scalar product of  
\begin{equation}
{\bf T}=\left( \begin{array}{c}
          T^a_{\sigma_a' \sigma_a"}\\
          T^b_{\sigma_b',\sigma_b"}\\
          T^c_{\sigma_c',\sigma_c"}\end{array}\right)
=\left(\begin{array}{c}
Abs[S( a\sigma', a\sigma"| p\pm \frac{q}{2} )_L+S( a\sigma', a\sigma"| p\pm\frac{q}{2} )_R]/2\\
Abs[S( b\sigma_b', b\sigma"| p\pm\frac{q}{2} )_L+S( b\sigma_b', b\sigma"| p\pm\frac{q}{2} )_R]/2\\
Abs[S( c\sigma_c', c\sigma"| p\pm\frac{q}{2} )_L+S( c\sigma_c', c\sigma"| p\pm\frac{q}{2} )_R]/2\end{array}\right)
\end{equation}
where $\sigma' \sigma"$ etc are $++, +-, -+$ and $--$, and
\begin{equation}
{\bf S}=\left(\begin{array}{c}
S^{bc}(\sigma_b\sigma_c,\sigma_b'\sigma_c")\\
S^{ca}(\sigma_c\sigma_a,\sigma_c'\sigma_a")\\
S^{ab}(\sigma_a\sigma_b,\sigma_a'\sigma_b")\end{array}\right)
=\left(\begin{array}{c}
S(b\sigma_b, b\sigma_b'|p+\frac{q}{2})_L \times S(c\sigma_c, c\sigma_c"|p+\frac{q}{2})_R^\dagger \\
S(c\sigma_c, c\sigma_c'|p+\frac{q}{2})_L \times S(a\sigma_a, a\sigma_a"|p+\frac{q}{2})_R^\dagger\\
S(a\sigma_a, a\sigma_a'|p+\frac{q}{2})_L \times S(b\sigma_b, b\sigma_b"|p+\frac{q}{2})_R^\dagger\end{array}\right)
\end{equation}

\begin{eqnarray}
N=\langle \sigma_a^L \sigma_b^L \sigma_c^R|T\cdot S|\sigma_a^R \sigma_b^R \sigma_c^L\rangle \nonumber
\end{eqnarray}
where $\sigma_a \sigma_b \sigma_c$ are $++-, +-+$ or $-++$ and according to whether the flavor of the quarks with  helicity $++$ coincides or not, the normalization of the three quark state is $\displaystyle \frac{2}{\sqrt{18}}$ or
$\displaystyle -\frac{1}{\sqrt{18}}$.
 \begin{table}
\begin{tabular}{c|cccccccc}
\hline\hline
 & $c_{uuaa}$ & $c_{uua}$ & $c_{uubc}$ & $c_{pp}$ & $c_{pu}$ & $c_{pua}$ & $c_{up}$ & $c_{upa}$ \\
\hline
$\mu=1$ & 4/18 & 1/18 & -2/18 & 1/18 & 1/18 & -2/18 & 1/18 & -2/18\\
$\mu=2$ & 1/18 & 4/18 & -2/18 & 1/18 & 1/18 & -2/18 & 1/18 & -2/18\\
\hline\hline
\end{tabular}
\end{table}

The three-body vertex matrix elements $\displaystyle G^{NVN}(p\pm \frac{q}{2})$ is calculated as follows.
I define the vertex matrix element on the left-handed quark 'a' and spectator quarks 'b' and 'c' as $QxyLXz$ where $xy$ specify whether the helicity of spectators in the initial state and the final state respectively are parallel(p) or unparallel(u),  $z$ defines the type of helicity flip of the spectators and $X$ defines the type of helicity flip of the vertex. The corresponding two body matrix element is defined as $NxyLXz$. In the calculation of left-handed quark contribution in $\Lambda_{\sigma_a'\sigma_a"}$, I measure 
\begin{equation}
Q=\langle \sigma_a^L \sigma_b^L \sigma_c^R|S\Lambda S^\dagger\cdot S|\sigma_a^R \sigma_b^R \sigma_c^L\rangle  
\end{equation}
where
\begin{equation}
{\bf \Lambda}=\left(\begin{array}{c}
S_{\sigma_a^L \sigma_a'}\Lambda_{\sigma_a'\sigma_a"}S_{\sigma_a^R\sigma_a"}^\dagger\\
S_{\sigma_b^L \sigma_b'}\Lambda_{\sigma_b'\sigma_b"}S_{\sigma_b^R\sigma_b"}^\dagger\\
S_{\sigma_c^L \sigma_c'}\Lambda_{\sigma_c'\sigma_c"}S_{\sigma_c^R\sigma_c"}^\dagger
\end{array}\right)
=\left(\begin{array}{c}
S(a\sigma_a, a\sigma_a'|p+\frac{q}{2})_L \Lambda_{\sigma_a'\sigma_a"} S(a\sigma_a, a\sigma_a"|p+\frac{q}{2})_R^\dagger \\
S(b\sigma_b, b\sigma_b'|p+\frac{q}{2})_L \Lambda_{\sigma_b'\sigma_b"} S(b\sigma_b, b\sigma_b"|p+\frac{q}{2})_R^\dagger\\
S(c\sigma_c, c\sigma_c'|p+\frac{q}{2})_L \Lambda_{\sigma_c'\sigma_c"} S(c\sigma_c, c\sigma_c"|p+\frac{q}{2})_R^\dagger\end{array}\right)
\end{equation}
as a sum of the matrix elements given in the Appendix.

I define 
\begin{eqnarray}
&&\Lambda ppL1=Re[QppL1]/Re[NppL1], \quad \Lambda ppL2=Re[QppL2]/Re[NppL2]  ,\nonumber\\
&&\Lambda ppL3=Re[QppL3]/Re[NppL3] , \quad \Lambda ppL4=Re[QppL4]/Re[NppL4] \nonumber\\
&&\Lambda puL1=\frac{Re[c_{pu}QpuL1+c_{pua}QpuL1a]}{Re[c_{pu}NpuL1+c_{pua}NpuL1a]} , \quad \Lambda puL2=\frac{Re[c_{pu}QpuL2+c_{pua}QpuL2a]}{Re[c_{pu}NpuL2+c_{pua}NpuL2a]} , \nonumber\\
&&\Lambda puL3=\frac{Re[c_{pu}QpuL3+c_{pua}QpuL3a]}{Re[c_{pu}NpuL3+c_{pua}NpuL3a]}, \quad \Lambda puL4=\frac{Re[c_{pu}QpuL4+c_{pua}QpuL4a]}{Re[c_{pu}NpuL4+c_{pua}NpuL4a]} \nonumber\\
&&\Lambda upL1=\frac{Re[c_{up}QupL1+c_{upa}QupL1a]}{Re[c_{up}NupL1+c_{upa}NupL1a]}, \quad \Lambda upL2=\frac{Re[c_{up}QupL2+c_{upa}QupL2a]}{Re[c_{up}NupL2+c_{upa}NupL2a]}, \nonumber\\
&&\Lambda upL3=\frac{Re[c_{up}QupL3+c_{upa}QupL3a]}{Re[c_{up}NupL3+c_{upa}NupL3a]} , \quad \Lambda upL4=\frac{Re[c_{up}QupL4+c_{upa}QupL4a]}{Re[c_{up}NupL4+c_{upa}NupL4a]}  \nonumber\\
&&\Lambda uuL1=\frac{Re[c_{uua}QuuL1a+c_{uuaa}QuuL1aa+c_{uubc}(QuuL1b+QuuL1c)]}
{Re[c_{uua}NuuL1a+c_{uuaa}NuuL1aa+c_{uubc}(NuuL1b+NuuL1c)]}\\
&&\Lambda uuL2=\frac{Re[c_{uua}QuuL2a+c_{uuaa}QuuL2aa+c_{uubc}(QuuL2b+QuuL2c)]}{Re[c_{uua}NuuL2a+c_{uuaa}NuuL2aa+c_{uubc}(NuuL2b+NuuL2c)]} , \nonumber\\
&&\Lambda uuL3=\frac{Re[c_{uua}QuuL3a+c_{uuaa}QuuL3aa+c_{uubc}(QuuL3b+QuuL3c)]}{Re[c_{uua}NuuL3a+c_{uuaa}NuuL3aa+c_{uubc}(NuuL3b+NuuL3c)]},\\
&&\Lambda uuL4=\frac{Re[c_{uua}QuuL4a+c_{uuaa}QuuL4aa+c_{uubc}(QuuL4b+QuuL4c)]}{Re[c_{uua}NuuL4a+c_{uuaa}NuuL4aa+c_{uubc}(NuuL4b+NuuL4c)]} \nonumber
\end{eqnarray}
and diagonalize the left-handed quark contribution
\[
\left(
\begin{array}{cc}
 \Lambda ppL1+\Lambda puL1+\Lambda upL1+\Lambda uuL1 & \Lambda ppL4+\Lambda puL4+\Lambda upL4+\Lambda uuL4\\
 \Lambda ppL3+\Lambda puL3+\Lambda upL3+\Lambda uuL3 & \Lambda ppL2+\Lambda puL2+\Lambda upL2+\Lambda uuL2 
\end{array}
\right),
\]
take eigenvalues and multiply the normalization factor $18^2$.
I calculate also the corresponding matrix elements of right-handed fermions, and add the two.

Number of samples is 52 and the error estimation was done in by using Bootstrap resampling command of Mathematica with 5000 iterations. (I calculate standard deviation of left-handed quark contribution divided by the mean of the left-handed quark contribution and the corresponding value of the right-handed quark contribution and multiplied the mean to the sum of the left-handed quark contribution and the right-handed quark contribution.)

\begin{figure}
\begin{center}
\includegraphics[width=6cm,angle=0,clip]{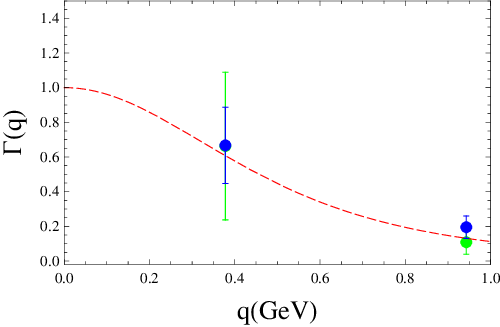}
\caption{The charge form factor of a proton using the DWF $m_u=0.01/a$ gauge fixed such that the fermion on the left wall and the right wall are correlated by self-dual gauge field, and the dipole fit with $M^2=(0.71$GeV$)^2$. (color online)}
\label{charge_plt_2}
\end{center}
\end{figure}

In contrast to $q=0$ case, the fluctuation is relatively large, as shown in the Fig.\ref{charge_plt_2}.
The contribution of $\Lambda_{V_1}$(blue (dark)) and $\Lambda_{V_2}$(green) are consistent with a dipole fit of $M^2=(0.71$GeV$)^2$ within errors. The standard deviation of $\Lambda_{V_2}$ is larger than that of $\Lambda_{V_1}$.

The nucleon electromagnetic form factor of the domain wall fermion was calculated recently by LHPC collaboration \citep{LHPC09}. The electric form factor $G_E(q^2)$ in their simulation has a small pion mass dependence, but the momentum dependence is flatter and the absolute value is larger than the phenomenological dipole fit. 
I guess the main difference from this calculation comes from the topology of the manifold. In our simulation, the helicity $h_1h_2$ of $\lambda_{h_1h_2}$ and $T_{h_1h_2}$ are the same, but when the boundary condition of the three quarks of the vertex $QxyLXz$ and that of the normalization $NxyLXz$ are fixed to be the same, $h_1h_2$ of the two are not necessarily the same and in this case the momentum dependence becomes flatter. 

\section{Discussion and Conclusion}
I showed that the discrepancy on the critical flavor number of fermions $N_f^c$ for the appearance of the infrared fixed point  derived from the Schr\"odinger functional method and the Bjorken sum rule and GDH sum rule and our lattice simulation could be resolved by assuming the topological structure of the infrared fixed point is not that of $U(1)^3$ but that of $G_2$ with triality automorphism.

The Lie algebra $g_2$ has the 8 dimensional subspace $g_0$ which is isomorphic to $sl_3(C)$ and the 3 dimensional subspace $W$ and $W^*$.  $sl_3(C)$ is not 
isomorphic to $su(3)$, but by imposing quaternion real condition, the subspace
isomorphic to $su(3)$ could be selected.

The lattice simulation of QCD running coupling from quark-gluon coupling in
Coulomb gauge suggests infrared fixed point of $\alpha_s(0)\sim \pi$ which is
predicted in the BLM renormalization scheme.  The strong interaction sum rules obtained in the infinite momentum frame with holographic variable $z$ in addition to the 3+1 dimensional space time is 5 dimensional and similar to ours.
I expect the similarity originates from the similar topological structure of the manifold, i.e. ${\bf S}^3\times {\bf R}$ of the quaternion.

Effects of instanton on ${\bf S}^3\times{\bf R}$ topology in Coulomb gauge was studied in the quaternion bases\citep{vBH92}. I observed that the self-dual gauge field (instanton) plays important roles in the stability of the infrared QCD.

Using the domain wall fermion in quaternion bases, and taking into account that the quadratic term of a quaternion makes an octonion, I remarked that there are 3,6 and 5 dimensional stable manifolds and the exceptional Lie algebra  $g_2$ is the automorphism in this space. The $g_2$ has triality automorphism and the topology of the infrared fixed point is not trivial.
I extended  \citep{sf09b} and \citep{sf09c} to a larger number of gauge configurations, and calculated the form factor at $q=0$ of the proton choosing a specific gauge that the fermion on the left domain wall and the right domain wall are correlated by the instanton-like self-dual gauge fields. I observed also via Ward identity through the difference of the inverse propagator at momentum p+q/2 and at p-q/2, charge form factor of the proton at $q\sim 0.4$GeV and $0.9$GeV agrees with the phenomenological dipole fit. Our simulation differ from the standard lattice simulation due to different boundary condition on three quarks. 

The triality automorphism gives another possible solution to the $U(1)$ problem, i.e. the $u_L-, d_L-$ and $s_L-$quarks need not have the same phase $e^{-i\theta}$ and the $u_R-, d_R-$ and $s_R-$quarks the same phase $e^{i\theta}$ due to extra degrees of freedom. It could affect or change the interpretation of the width of $\eta\to\pi^+\pi^0\pi^-$ decay,  which is about factor three larger than the standard current algebra result\citep{RT81}.

I used information of the color diagonal matrix elements only of quark propagators, and ignored the vector product part of quaternions\citep{KR91}.   I need to calculate other form factors of a proton and check the correlation of the three quarks in a proton. It is necessary to extend the simulation to a larger lattice and calculate form factors at more momentum points and examine the triality automorphisms in more detail. 
\vskip 0.5 true cm
I would like to thank Dr.Francesco Sannino for valuable information on the anomaly matching, Professor Stan Brodsky for attracting attention to the ref. \citep{BS08} and helpful discussion, and Dr. Alex Deur for sending me the experimental data.
I acknowledge helpful suggestion of paying attention on recent DSE approach \citep{FMP09} and on the Pinch Technique \citep{BP09}, from one of the referees of a journal. 

The numerical simulation was performed on Hitachi-SR11000 at High Energy Accelerator Research Organization(KEK) under a support of its Large Scale Simulation Program (No.07-04 and No.08-01), and on NEC-SX8 at Yukawa institute of theoretical physics of Kyoto University.
\newpage
\appendix
\subsection{$\gamma$ matrices and triality}
I use the $\gamma$ matrices, as follows:
\begin{equation}
\gamma_k=\left(\begin{array}{cc}
0 & i\sigma_k \\
-i\sigma_k & 0\end{array}\right)
\qquad
\gamma_5=\left(\begin{array}{cccc}
1 & 0 & 0 & 0\\
0 & 1 & 0 & 0\\
0 & 0 &-1 & 0\\
0 & 0 & 0 & -1
\end{array}\right),
\qquad
\gamma_4=\left(\begin{array}{cccc}\
0 & 0 & 1 & 0\\
0 & 0 & 0 & 1\\
1 & 0 & 0 & 0\\
0 & 1 & 0 & 0
\end{array}\right)\nonumber
\end{equation}

\begin{equation}
\gamma_1=\left(\begin{array}{cccc}
0 & 0 & 0 & i\\
0 & 0 & i & 0\\
0 & -i & 0 & 0\\
-i & 0 & 0 & 0
\end{array}\right),
\qquad
\gamma_2=\left(\begin{array}{cccc}
0 & 0 & 0 & -1\\
0 & 0 & 1 & 0\\
0 & 1 & 0 & 0\\
-1 & 0 & 0 & 0
\end{array}\right),
\qquad
\gamma_3=\left(\begin{array}{cccc}
0 & 0 & i & 0\\
0 & 0 & 0 & -i\\
-i & 0 & 0 & 0\\
0 & i & 0 & 0
\end{array}\right),\nonumber
\end{equation}

\begin{equation}
\gamma_4\gamma_2\gamma_5=\left(\begin{array}{cccc}
0 & 1 & 0 & 0\\
-1 & 0 & 0 & 0\\
0 & 0 & 0 & 1\\
0 & 0 & -1 & 0
\end{array}\right)
\qquad
\gamma_4\gamma_2=\left(\begin{array}{cccc}
0 & 1 & 0 & 0\\
-1 & 0 & 0 & 0\\
0 & 0 & 0 & -1\\
0 & 0 & 1 & 0
\end{array}\right)\nonumber
\end{equation}
\begin{equation}
\gamma_4\gamma_1\gamma_5=\left(\begin{array}{cccc}
0 & -i & 0 & 0\\
-i & 0 & 0 & 0\\
0 & 0 & 0 & -i\\
0 & 0 & -i & 0
\end{array}\right)
\qquad
\gamma_4\gamma_3\gamma_5=\left(\begin{array}{cccc}
-i & 0 & 0 & 0\\
0 & i & 0 & 0\\
0 & 0 & -i & 0\\
0 & 0 & 0 & i
\end{array}\right)\nonumber
\end{equation}
\newpage
\subsection{The matrix elements of the triality transformation of the $g_2$ algebra.}

In $G_{23}$, I replaced $\xi_3 \to \xi_{34}$ to $\xi_3\to -\xi_{34}$, 
and in $G_{12}$, $\xi_4\to -\xi_{1234}$ to $\xi_4\to -\xi_{123}$ and $x^4\to-\xi_0$
to $x^4\to \xi_0$ written in \citep{Cartan66} (probably typos). 
$G_{13}G_{12}$ does not agree with $G_{(123)}$ of \citep{Cartan66}. I think
$\xi_4\to -x^{4'}$ there should be replaced by $\xi_4\to x^4$ and I define
${^t(G}_{12}G_{13})\equiv {G}_{123}$ and ${^t(G}_{13}G_{12})\equiv{G}_{132}$,  which correspond
 to $G_{(123)}$ and $G_{(132)}$ respectively.
 Under this triality automorphism, the set $\{F,\Phi,\Psi\}$ is invariant \`a signe pr\`es. (In the tables $\bar 1$ indicates -1.)
\vskip 0.5 true cm

\begin{figure}
\begin{center}
\includegraphics[width=17cm,angle=0,clip]{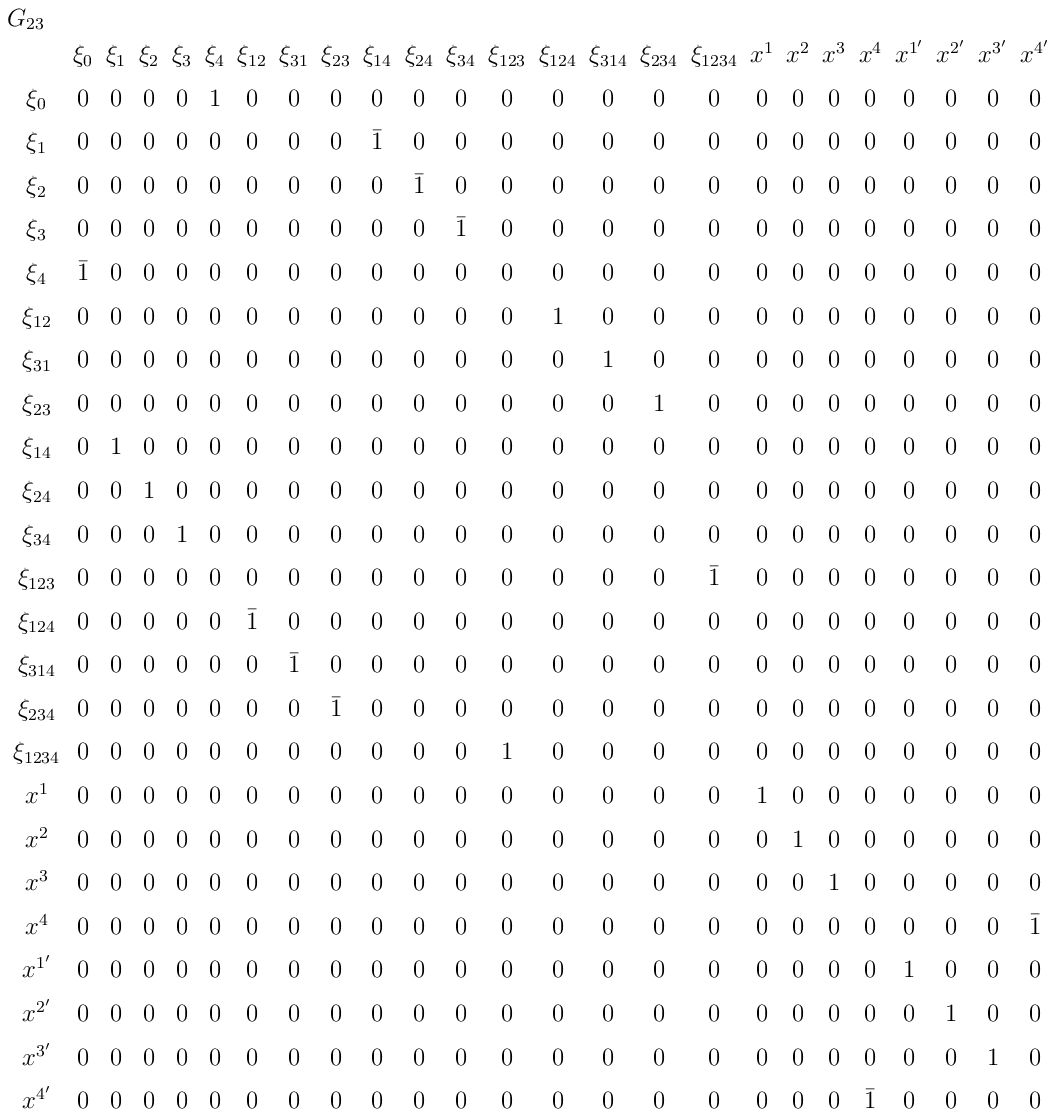}
\label{G23}
\end{center}
\end{figure}
\pagestyle{empty}
\begin{figure}
\begin{center}
\includegraphics[width=17cm,angle=0,clip]{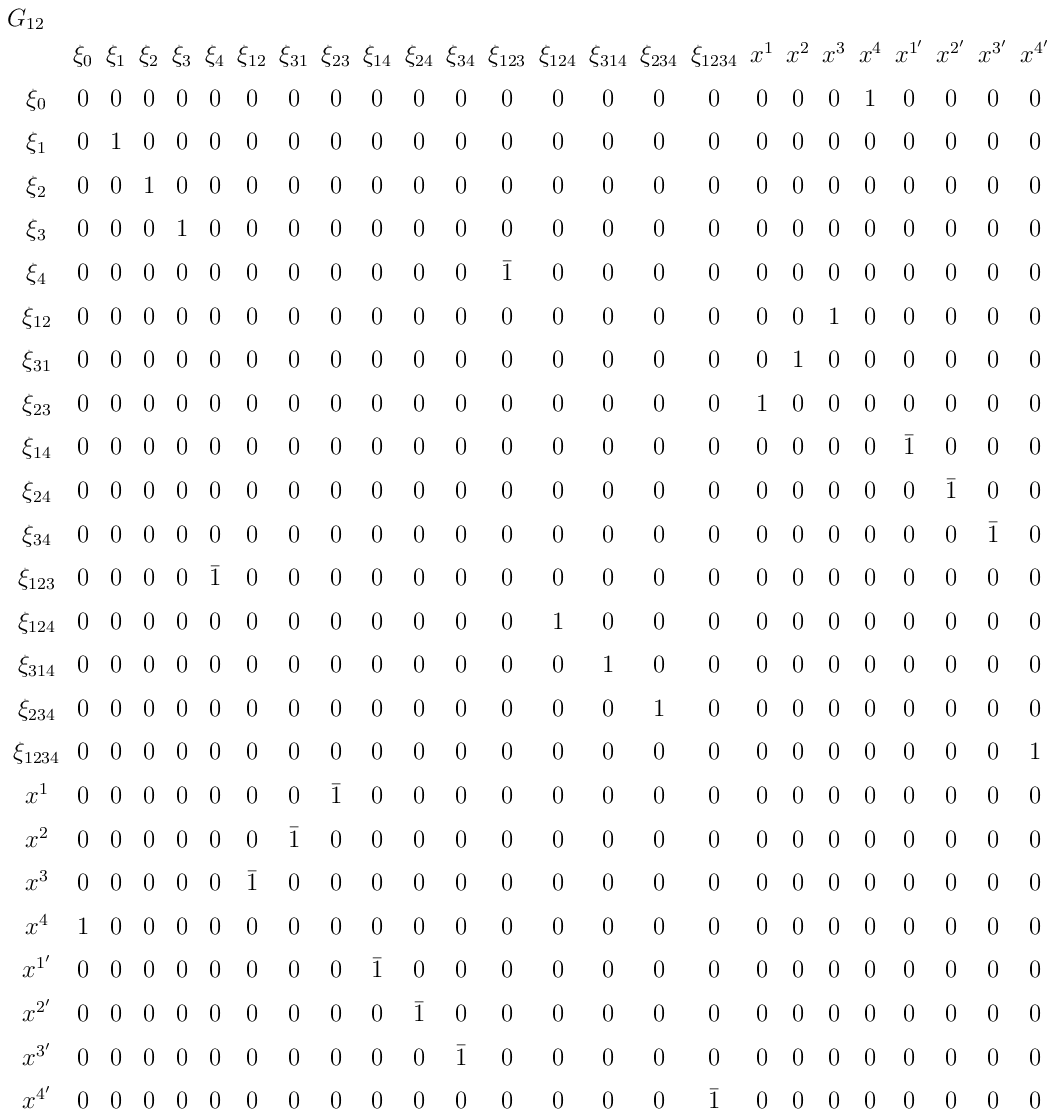}
\label{G23}
\end{center}
\end{figure}
\pagestyle{empty}
\begin{figure}
\begin{center}
\includegraphics[width=17cm,angle=0,clip]{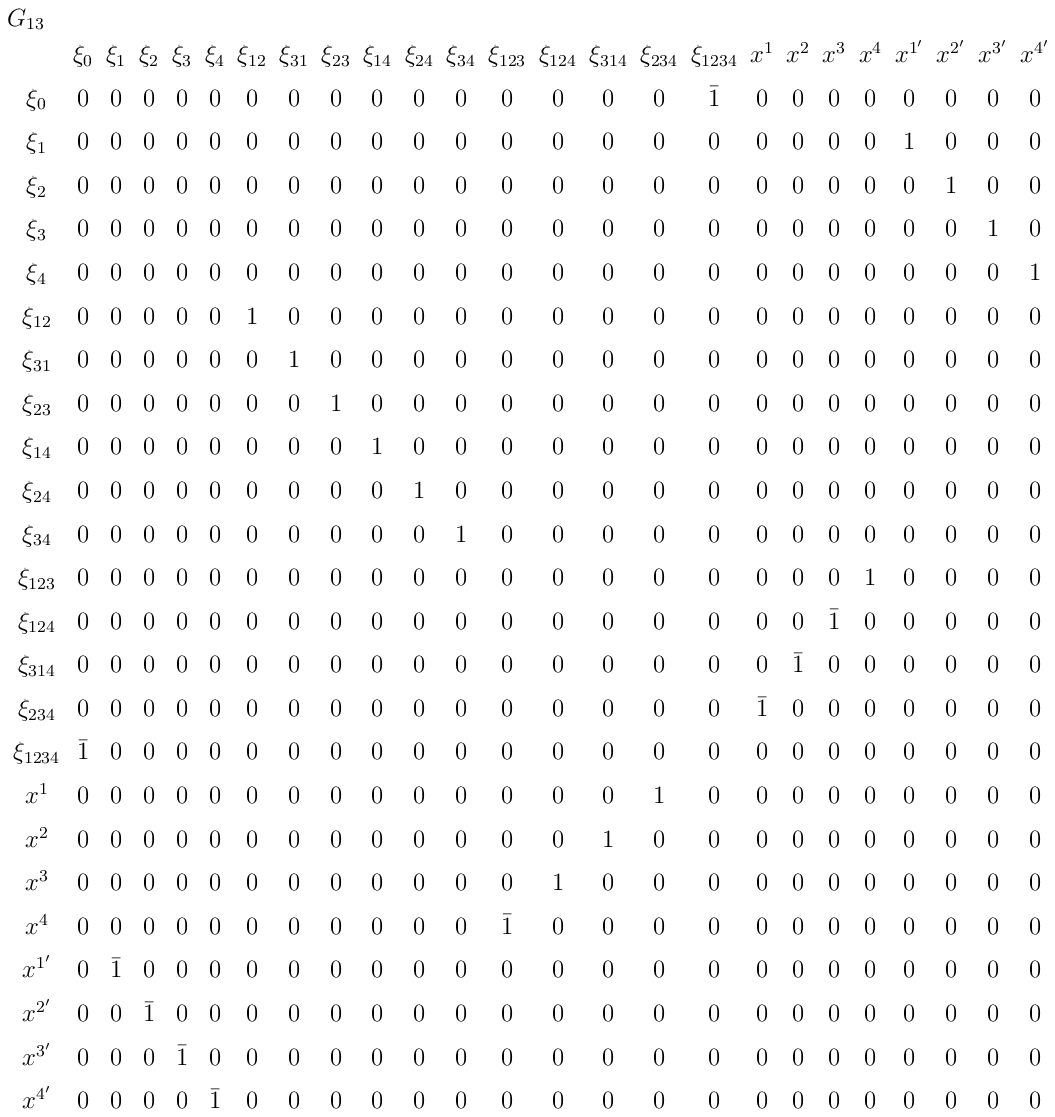}
\label{G13}
\end{center}
\end{figure}
\begin{figure}
\begin{center}
\includegraphics[width=17cm,angle=0,clip]{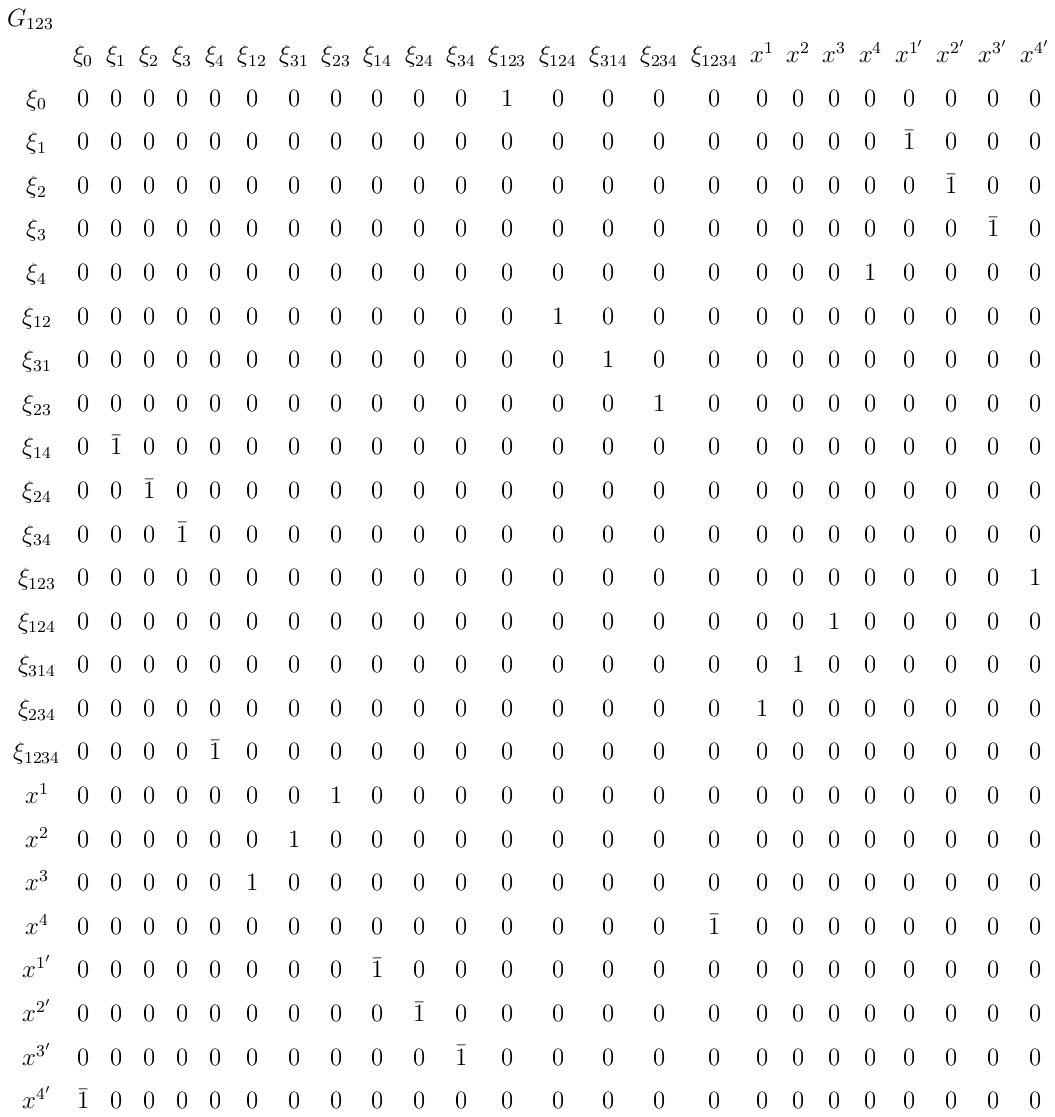}
\label{G123}
\end{center}
\end{figure}
\begin{figure}
\begin{center}
\includegraphics[width=17cm,angle=0,clip]{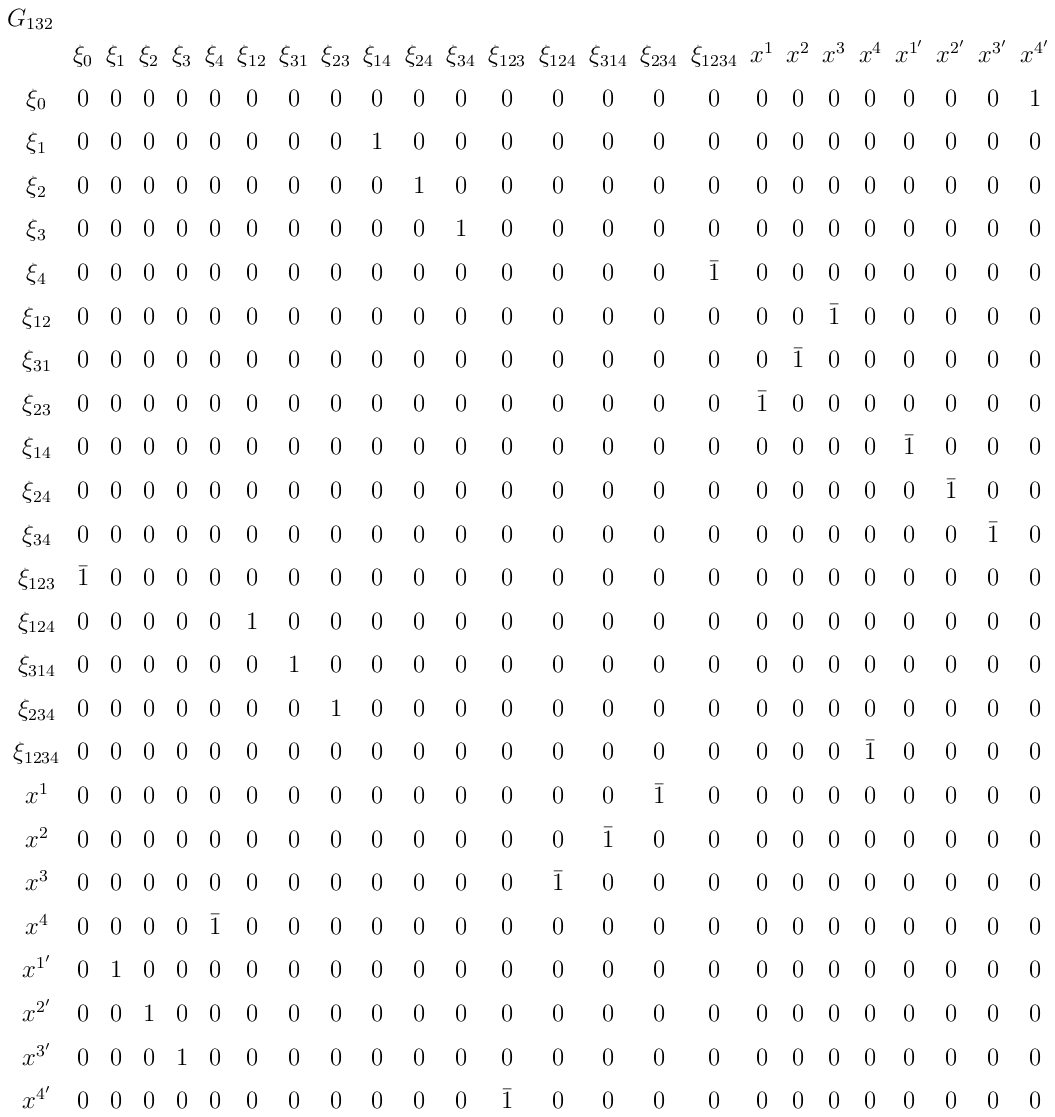}
\label{G132}
\end{center}
\end{figure}

\newpage
\subsection{The matrix elements of the vertex $QxyLXz$ and the normalization $NxyLXz$ in the form factor calculation}

In the calculation of the $G_a^{NVN}$ with $\Lambda_{\mu=1}$, I used the martix elements of the vertex $QxyLXz$ and the normalization $NxyLXz$ as follows and the corresponding matrix elements of L and R interchanged, i.e. $QxyRXz$ and $NxyRXz$. The vertices of $G_b^{NVN}$ with $\Lambda_{\mu=2}$ is also calculated similarly.

$QuuL1a={_L\langle} + + -|S_{++}\Lambda_{++}S_{++}^\dagger\cdot S_{+-+-}|+ + -\rangle_R, \\ \qquad  NuuL1a={_L\langle}T_{++}\cdot S_{+-+-}\rangle_R$, 

$QuuL1aa={_L\langle} + - +|S_{++}\Lambda_{++}S_{++}^\dagger\cdot S_{-+-+}|+ - +\rangle_R, \\ \qquad NuuL1aa={_L\langle}T_{++}\cdot S_{-+-+}\rangle_R$,

$QuuL1b={_L\langle} + + -|S_{++}\Lambda_{++}S_{++}^\dagger\cdot S_{+-+-}|+ - +\rangle_R, \\ \qquad NuuL1b={_L\langle}T_{++}\cdot S_{+-+-}\rangle_R$, 

$QuuL1c={_L\langle} + - +|S_{++}\Lambda_{++}S_{++}^\dagger\cdot S_{-+-+}|+ - +\rangle_R, \\ \qquad NuuL1c={_L\langle}T_{++}\cdot S_{-+-+}\rangle_R$,
 
$QuuL2a={_L\langle} + - +|S_{+-}\Lambda_{--}S_{-+}^\dagger\cdot S_{-+-+}|+ - +\rangle_R, \\ \qquad NuuL2a={_L\langle}T_{--}\cdot S_{-+-+}\rangle_R$,

$QuuL2aa={_L\langle} + + -|S_{+-}\Lambda_{--}S_{-+}^\dagger\cdot S_{+-+-}|+ + -\rangle_R, \\ \qquad NuuL2aa={_L\langle} T_{--}\cdot S_{+-+-}\rangle_R$,

$QuuL2b={_L\langle} + + -|S_{+-}\Lambda_{--}S_{-+}^\dagger\cdot S_{+--+}|+ - +\rangle_R, \\ \qquad  NuuL2b={_L\langle}T_{--}\cdot S_{+--+}\rangle_R$,

$QuuL2c={_L\langle} + - +|S_{+-}\Lambda_{--}S_{-+}^\dagger\cdot S_{+-+-}|+ + -\rangle_R, \\ \qquad  NuuL2c={_L\langle} T_{--}\cdot S_{+-+-}\rangle_R$,

$QuuL3a={_L\langle} + - +|S_{++}\Lambda_{+-}S_{-+}^\dagger\cdot S_{-+-+}|+ - +\rangle_R, \\ \qquad  NuuL3a={_L\langle}T_{+-}\cdot S_{-+-+}\rangle_R$,

$QuuL3aa={_L\langle} + + -|S_{++}\Lambda_{+-}S_{-+}^\dagger\cdot S_{+-+-}|+ + -\rangle_R, \\ \qquad  NuuL3a={_L\langle}T_{+-}\cdot S_{+-+-}\rangle_R $,

$QuuL3b={_L\langle} + + -|S_{++}\Lambda_{+-}S_{-+}^\dagger\cdot S_{+--+}|+ - +\rangle_R,\\ \qquad  NuuL3b={_L\langle}T_{+-}\cdot S_{+--+}\rangle_R$,

$QuuL3c={_L\langle} + - +|S_{++}\Lambda_{+-}S_{-+}^\dagger\cdot S_{-++-}|+ + -\rangle_R, \\ \qquad  NuuL3c={_L\langle}T_{+-}\cdot S_{-++-}\rangle_R$,

$QuuL4a={_L\langle} + - +|S_{++}\Lambda_{-+}S_{-+}^\dagger\cdot S_{-+-+}|+ - +\rangle_R, \\ \qquad  NuuL4a={_L\langle}T_{-+}\cdot S_{-+-+}\rangle_R$,

$QuuL4aa={_L\langle} + + -|S_{+-}\Lambda_{-+}S_{++}^\dagger\cdot S_{+--+}|+ - +\rangle_R, \\ \qquad  NuuL4aa={_L\langle}T_{-+}\cdot S_{+--+}\rangle_R$,

$QuuL4b={_L\langle} + + -|S_{+-}\Lambda_{-+}S_{++}^\dagger\cdot S_{+--+}|+ - +\rangle_R, \\ \qquad  NuuL4b={_L\langle}T_{-+}\cdot S_{+--+}\rangle_R$,

$QuuL4c={_L\langle} + - +|S_{+-}\Lambda_{-+}S_{++}^\dagger\cdot S_{-+-+}|+ + -\rangle_R, \\ \qquad  NuuL4c={_L\langle}T_{-+}\cdot S_{-+-+}\rangle_R$,

$QpuL1={_L\langle} - + +|S_{-+}\Lambda_{++}S_{++}^\dagger\cdot S_{+++-}|+ + -\rangle_R, \\ \qquad  NpuL1={_L\langle}T_{++}\cdot S_{+++-}\rangle_R$, 

$QpuL1a={_L\langle} - + +|S_{-+}\Lambda_{++}S_{++}^\dagger\cdot S_{++-+}|+ - +\rangle_R, \\ \qquad  NpuL1a={_L\langle}T_{++}\cdot S_{++-+}\rangle_R$,   

$QpuL2={_L\langle} - + +|S_{--}\Lambda_{--}S_{-+}^\dagger\cdot S_{+++-}|+ + -\rangle_R, \\ \qquad  NpuL2={_L\langle}T_{--}\cdot S_{+++-}\rangle_R$, 

$QpuL2a={_L\langle} - + +|S_{--}\Lambda_{--}S_{-+}^\dagger\cdot S_{++-+}|+ - +\rangle_R, \\ \qquad  NpuL2a={_L\langle}T_{--}\cdot S_{++-+}\rangle_R$, 

$QpuL3={_L\langle} - + +|S_{-+}\Lambda_{+-}S_{-+}^\dagger\cdot S_{+++-}|+ + - \rangle_R, \\ \qquad  NpuL3={_L\langle}T_{+-}\cdot S_{+++-}\rangle_R$,  

$QpuL3a={_L\langle} - + +|S_{-+}\Lambda_{+-}S_{--}^\dagger\cdot S_{++-+}|+ - +\rangle_R, \\ \qquad  NpuL3a={_L\langle}T_{+-}\cdot S_{++-+}\rangle_R$,

$QpuL4={_L\langle} - + +|S_{--}\Lambda_{-+}S_{++}^\dagger\cdot S_{+++-}|+ + - \rangle_R, \\ \qquad  NpuL4={_L\langle}T_{-+}\cdot S_{+++-}\rangle_R$,  

$QpuL4a={_L\langle} - + +|S_{--}\Lambda_{-+}S_{++}^\dagger\cdot S_{++-+}|+ - +\rangle_R, \\ \qquad  NpuL4a={_L\langle}T_{-+}\cdot S_{++-+}\rangle_R$,

$QupL1={_L\langle} + + -|S_{++}\Lambda_{++}S_{+-}^\dagger\cdot S_{+-++}|- + +\rangle_R, \\ \qquad  NpuL1={_L\langle}T_{++}\cdot S_{+-++}\rangle_R$, 

$QupL1a={_L\langle} + - +|S_{++}\Lambda_{++}S_{+-}^\dagger\cdot S_{-+++}|- + +\rangle_R, \\ \qquad  NupL1a={_L\langle}T_{++}\cdot S_{-+++}\rangle_R$, 

$QupL2={_L\langle} + + -|S_{+-}\Lambda_{--}S_{--}^\dagger\cdot S_{+-++}|- + +\rangle_R,\\ \qquad  NupL2={_L\langle}T_{--}\cdot S_{+-++}\rangle_R$, 

$QupL2a={_L\langle} + - +|S_{+-}\Lambda_{--}S_{--}^\dagger\cdot S_{-+++}|- + +\rangle_R, \\ \qquad  NupL2a={_L\langle} T_{--}\cdot S_{-+++}\rangle_R$, 

$QupL3={_L\langle} + + -|S_{++}\Lambda_{+-}S_{--}^\dagger\cdot S_{+-++}|- + +\rangle_R, \\ \qquad  NupL3={_L\langle} T_{+-}\cdot S_{+-++}\rangle_R$, 

$QupL3a={_L\langle} + - +|S_{++}\Lambda_{+-}S_{--}^\dagger\cdot S_{-+++}|- + +\rangle_R, \\ \qquad  NupL3a={_L\langle} T_{+-}\cdot S_{-+++}\rangle_R$, 

$QupL4={_L\langle} + + -|S_{+-}\Lambda_{-+}S_{+-}^\dagger\cdot S_{+-++}|- + +\rangle_R, \\ \qquad  NupL4={_L\langle}T_{-+}\cdot S_{+-++}\rangle_R$, 

$QupL4a={_L\langle} + - +|S_{+-}\Lambda_{-+}S_{+-}^\dagger\cdot S_{-+++}|- + +\rangle_R, \\ \qquad  NupL4a={_L\langle}T_{-+}\cdot S_{-+++}\rangle_R$,

$QppL1={_L\langle} - + +|S_{-+}\Lambda_{++}S_{+-}^\dagger\cdot S_{++++}|- + +\rangle_R, \\ \qquad  NppL1={_L\langle}T_{++}\cdot S_{++++}\rangle_R$, 

$QppL2={_L\langle} - + +|S_{--}\Lambda_{--}S_{--}^\dagger\cdot S_{++++}|- + +\rangle_R, \\ \qquad  NppL2={_L\langle} T_{--}\cdot S_{++++}\rangle_R$, 

$QppL3={_L\langle} - + +|S_{-+}\Lambda_{+-}S_{--}^\dagger\cdot S_{++++}|- + +\rangle_R, \\ \qquad  NppL3={_L\langle} T_{+-}\cdot S_{++++}\rangle_R$, 

$QppL4={_L\langle} - + +|S_{--}\Lambda_{-+}S_{+-}^\dagger\cdot S_{++++}|- + +\rangle_R, \\ \qquad  NppL4={_L\langle} T_{-+}\cdot S_{++++}\rangle_R$.

\bibliographystyle{plainnat}

\end{document}